\documentclass[seceq]{ptptex}

\usepackage{graphicx}
\usepackage{wrapft}

\notypesetlogo                   

\title{
Cluster Abundance Evolution and the Sunyaev-Zeldovich Effect
\\ in Various Cosmological Models
}

\author{
Kenji \textsc{Tomita}\footnote{E-mail: tomita@yukawa.kyoto-u.ac.jp}
}

\inst{
Yukawa Institute for Theoretical Physics, Kyoto University, \\
Kyoto 606-8502, Japan
}

\recdate{June 1, 2003}

\abst{
The redshift dependence of the observed cluster abundance due to the 
Sunyaev-Zeldovich effect (SZE) is studied in various cosmological models, including 
flat and open homogeneous (CDM) models and an inhomogeneous model with 
a large-scale local void. The Press-Schechter formalism is used to derive the 
abundance at epochs in the range $0 < z < 2$, and the cluster mass 
limit $M_{\rm lim}$ is obtained from a flux limit for SZE. It is shown 
that SZE is useful for constraining 
the cosmological model parameters, and the abundance in the inhomogeneous model
 may be comparable with that in the low-density homogeneous models. 
The significance of relative difference of abundances in the various models 
is discussed. 
}

\begin{document}

\maketitle

\section{Introduction}

The abundance of clusters has been extensively studied by many people 
in X-ray surveys to constrain the cosmological parameters: e.g., 
Bahcall and Fan,\cite{bahfan} Viana and Liddle,\cite{vian} Eke, Cole and Frenk,\cite{eke} 
and Kitayama and Suto.\cite{ks96a,ks96b,ks97}
The cluster abundance found in the submm survey based on the Sunyaev-Zeldovich 
effect (SZE) has also been studied by Haiman, Mohr and Holder,\cite{haiman}
Holder, Haiman and Mohr,\cite{hold00,hold01} Kitayama, Sasaki and
Suto,\cite{kita} and Fan and Chiueh.\cite{fanchiu} This SZE survey is
expected to be most important in clarifying 
the evolution of clusters and constraining the cosmological parameters.

In this paper we study the cluster abundance obtained in the SZE survey, whose 
observational conditions were set to correspond to the interferometric 
arrays in the AMIBA project. 
The cluster mass limit $M_{\rm lim}$ for deriving the 
abundance is determined using the expression for the flux $S_\nu$ given 
by Kitayama and Suto,\cite{ks96b} which is different from that used by Fan and 
Chiueh.\cite{fanchiu} This leads to a difference in the behavior of
the resulting cluster abundances from that reported in their paper.
 
In \S 2, we describe the formulation for deriving the cluster abundance 
in the SZE survey. In \S 3 we present the results in various cosmological
models. Here, we first consider four representative homogeneous 
cosmological models: LCDM with $(\Omega_0, \lambda_0) = (0.3, 0.7), \ 
h=0.7$ and $\Gamma = 0.25$, \ OCDM with $(0.3, 0), \ h=0.7$ and
$\Gamma = 0.25$, SCDM with $(1.0, 0), \ h=0.5$ and $\Gamma = 0.5$, and 
$\tau$CDM with with $(1.0, 0), \ h=0.5$ and $\Gamma = 0.25$, where 
$\Gamma$ is the CDM shape parameter and $H_0 = 100h$ km s$^{-1}$ Mpc$^{-1}$.

At present, most cosmological observations, including SDSS,\cite{zehavi} high-redshift 
supernovae\cite{sch,riessa,riessb,perl} and WMAP,\cite{bennett,sperg} 
support a flat homogeneous model with nonzero 
cosmological constant. However, the presently observed values of the Hubble 
constant seem to be non-uniform,\cite{keeton,courb,fassn,will,tada,kocha} 
though they include rather large uncertainties;
that is, the local median value seems to be larger than that of the global
median value by a factor approximately equal to $1.2$. If the non-uniformity of the
Hubble constant is found to be real, we may have to use inhomogeneous
models to describe the cosmological observations.
In the magnitude-redshift diagram of supernovae of type Ia, moreover,
the point representing the recent data\cite{new} with $z = 1.7$ deviates from
the curve predicted by the concordant model with nonzero cosmological constant.
If additional supernovae with $z > 1.5$ exhibiting a similar trend
are obtained, models different from the concordant model may be
needed in order to account for the data.  Taking this situation into 
consideration, we consider here an inhomogeneous model with a 
large-scale local void as a representative inhomogeneous model, in which
there are inner and outer homogeneous regions (I and II) and a 
spherical boundary.\cite{tma,tmb,tmc,tmd} 

The objectives motivating our study of this
inhomogeneous model are discussed in previous papers.\cite{tme,tmf}
The first of these objectives is the introduction of the inhomogeneity of the Hubble
constant, and the second is the explanation of the accelerating
behavior of SNIa in the absense of a cosmological constant.
The cosmological parameters in the two regions are 
$(\Omega_0^{\rm I}, \lambda_0^{\rm I}) =(0.3, 0), \ \Gamma^{\rm I} =0.25, 
\ H_0^{\rm I} = 70$ km s$^{-1}$ Mpc$^{-1}$ in the inner region (I) and 
$(\Omega_0^{\rm II}, \lambda_0^{\rm II}) =(1.0, 0), \ 
\Gamma^{\rm II} =0.5, \ H_0^{\rm II} = H_0^{\rm I} \times 0.82$ in the outer
 region (II). It is assumed that our observer is at the center for simplicity, 
and the spherical boundary corresponds to a redshift $z_1 = 0.067$. 
For the above set of parameters, the observed magnitude-redshift
relation of SNIa, including the above high redshift data, was reproduced in the
inhomogeneous model.\cite{tmd} 
The consistency of spherically symmetric inhomogeneous models with the
supernova data was recently discussed and examined also by Iguchi et al.\cite{iguchi}
In \S 3 we discuss the difference among the abundances in these
various models. In \S 4 we have concluding remarks.

\section{Number density and SZE}
\subsection{The observed number of clusters}
The comoving number density of clusters of mass $M$ with width $dM$ is
\begin{equation}
  \label{eq:g1}
n(M) dM = \Big({2 \over \pi}\Big)^{1/2} {\rho_0 \over M}{\delta_c(z)\over 
\sigma_0^2}{d\sigma_0 \over dM} \exp \Big(-{\delta_c^2 (z) 
\over 2\sigma_0^2} \Big),
\end{equation}
following the Press-Schechter formalism, where $\rho_0$ is the present mass 
density of the universe, \ $\delta_c(z)$ is the linear density threshold for 
collapse at redshift $z$, and $\sigma_0$ is the rms linear density 
perturbation on the scale corresponding to $M$. The expressions for 
$\delta_c (z)$ in homogeneous models are given in Kitayama and Suto's 
paper\cite{ks96b} (cf. their Appendix A). 

The differential number of clusters is expressed as
\begin{equation}
  \label{eq:g2}
{dN \over dz d\Omega} = {dV \over dz d\Omega} \int_{M_{\rm lim}} n(M) dM, 
\end{equation}
where $d\Omega$ is the solid angle element, $dV$ is the comoving volume 
element, and $M_{\rm lim}$ is the lower limit of the observed cluster mass, which 
is discussed in the next subsection. Here, the comoving volume element
$dV$ is given by
\begin{equation}
  \label{eq:g3}
dV = ct \ [d_{\rm A} (z)]^2 d\Omega (1 +z)^3,
\end{equation}
where $d_{\rm A}$ is the angular diameter distance. Following Viana and 
Liddle\cite{vian}, we assume $\sigma_0$ in the form
\begin{equation}
  \label{eq:g4}
\sigma_0 = \sigma_8 (z) \Big({R \over 8h^{-1}{\rm Mpc}}\Big)^{-\gamma (R)},
\end{equation}
where $\gamma (R) = (0.3 \Gamma +0.2) [2.92 + \log_{10} ({R /
8 h^{-1}{\rm Mpc}} ) ]$,
with $M \equiv {4 \over 3} \pi \rho_0 R^3$, and we have 
\begin{equation}
  \label{eq:g4a}
\sigma_8 (z) = \sigma_8 (0) g(\Omega(z),\lambda(z))/[(1+z)\ g(\Omega_0,\lambda_0)],
\end{equation}
where $g(\Omega,\lambda)$ is an approximate factor representing the growth of linear
density perturbations (Carroll, Press and Turner\cite{carroll}), given by 
\begin{equation}
  \label{eq:g4b}
g(\Omega,\lambda) = {5 \over 2}\Omega/[\Omega^{4/7}-\lambda+(1+\Omega/2)(1+\lambda/70)],
\end{equation}
and $\Omega(z)$ and $\lambda(z)$ are  $\Omega$ and $\lambda$ at the epoch $z$.
 
For homogeneous models the observed values of $\sigma_8 (0)$ have been derived by 
several groups: for instance, \  
$\sigma_8 (0) \Omega_0^{0.45} = 0.53\pm 0.05$ \ for \ $\lambda_0 = 0$ \ and \
 $\sigma_8 (0) \Omega_0^{0.53} = 0.53\pm 0.05$ \ for \ $\Omega_0+\lambda_0 =1$
by Pen\cite{pen} and Eke et al.\cite{eke}, \ $\sigma_8 (0) \Omega_0^{0.60} = 
0.50\pm 0.04 $ by Pierpaoli et al.,\cite{pierp}
\ $\sigma_8 (0) \Omega_0^{0.6} = 0.345\pm 0.05 $ \
by Fisher et al.\cite{fisher}, \ $\sigma_8 (0) \Omega_0^{0.5} = 0.33\pm 0.03 $ \
by Bahcall et al.\cite{bahc02}, \ $\sigma_8 (0) \Omega_0^{0.48 - 0.27\Omega_0}
 = 0.38$ \ by Viana, Nichol and Liddle\cite{viana02} \
and \ $\sigma_8 (0) \Omega_0^{0.5} = 0.48\pm 0.12 $ \
by Spergel et al.\cite{sperg} for the WMAP data. Here we adopt the following two sets of 
values:
\begin{equation}
  \label{eq:g5}
\sigma_8 (0) \Omega_0^p = 0.4 \ {\rm and} \ 0.5,
\end{equation}
where $p$ is \ $0.45$ \ and \ $ 0.53$ \ for models with \ $\lambda_0 = 0$ \ 
and flat models with \ $\Omega_0+\lambda_0 =1$, 
respectively.
 
\subsection{The lower mass limit $M_{\rm lim}$}
When photons pass through a cluster of hot electrons, a temperature 
decrease results and the black-body spectrum is distorted
due to inverse Compton scattering as
\begin{equation}
  \label{eq:g6}
\Delta T/T_{\rm CMB} = g(x) y,
\end{equation}
where the Compton $y$-parameter is
\begin{equation}
  \label{eq:g7}
y \equiv \int n_e \sigma_{\rm T} \Big({kT_{\rm gas} \over m_ec^2}\Big) dl,
\end{equation}
\begin{equation}
  \label{eq:g8}
g(x) \equiv {x \over \tanh (x/2)} -4.
\end{equation}
Here, $x \equiv h\nu/(kT_{\rm CMB}), where \nu$ is the CMB photon frequency, \ $n_e$ 
is the electron number density, \ $\sigma_{\rm T}$ is the Thomson cross 
section, \ $T_{\rm gas}$ is the temperature of the cluster gas, and the 
integration is along the line of sight. If $T_{\rm gas} \gg T_{\rm CMB}$, 
the flux of CMB photons changes from $S_{\nu}^{\rm CMB} \ (\equiv (2h\nu^3/c^2)
/(e^x -1))$ to
\begin{equation}
  \label{eq:g9}
S_\nu = S_\nu^{\rm CMB} {x e^x\over e^x -1} g(x) Y,
\end{equation}
where 
\begin{equation}
  \label{eq:g10}
Y = [d_{\rm A}]^{-2} \int y dA,
\end{equation}
with $dA$ the element of the projected area of a cluster.
For $\nu = 219$ GHz, we have $g(x) = 0$. For $\nu < 219$ 
GHz, $g(x) < 0$, and for $\nu > 219$ GHz, $g(x) > 0$
In the Array for Microwave Background (AMIBA) project 
(Fan and Chiueh\cite{fanchiu}), the parameters are $\nu = 90$ GHz and $x 
\approx 1.58$, which we use in the following. 
For an isothermal cluster with constant gas mass fraction, we
have
\begin{equation}
  \label{eq:g11}
Y = {\sigma_{\rm T} \over 2 m_e m_p} [d_{\rm A}]^{-2} 
f_{\rm ICM} (1+X) kT_{\rm gas} M,
\end{equation}
where $m_p$ is the photon mass, $X$ is the hydrogen mass
fraction, and $f_{\rm ICM}\equiv \Omega_{\rm B}/\Omega_0$ for
the present baryon density parameter $\Omega_{\rm B}$.
For an isothermal gas, $T_{\rm gas}$ is related to the
total cluster mass $M$ by
\begin{equation}
  \label{eq:g12}
kT_{\rm gas} = 5.2 \gamma (1+z) \Big({\rho_{\rm vir}(z) \over 
18\pi^2}\Big)^{1/3} \Big({M \over 10^{15} h^{-1} M_\odot}\Big)^{2/3}
\Omega_0^{1/3} {\rm kev},
\end{equation}
as shown by Kitamura, Sasaki and Suto,\cite{kita} and the expression 
for $\rho_{\rm vir}(z)$ (which is the ratio of the mean density of the 
virialized cluster
to the mean density of the universe at each epoch) is given 
in the paper of Kitamura and Suto\cite{ks96b} (cf. their Appendix A).
Here, $z$ is the redshift for cluster formation in principle, but
it is regarded here as the redshift at the epoch in which the 
cluster exists. Then, the total flux of the cluster is
\begin{equation}
  \label{eq:g13}
S_\nu = 25.5 h (1+z) \bar{g}(x) {1+X \over 2}\times
{\Omega_{\rm B} \over \Omega_0^{2/3}} \Big[{d_{\rm A}(z) \over
c/H_0}\Big]^{-2} \Big({\rho_{\rm vir}(z) \over 
18\pi^2}\Big)^{1/3} \Big({M \over 10^{15} h^{-1} M_\odot}\Big)^{5/3}
\ {\rm mJy},
\end{equation}
where $\bar{g}(x)$ is given by $\bar{g}(x) \equiv x^4 e^x
(e^x -1)^{-2} g(x)$. This expression is different from that given
in Fan and Chiueh's paper, based on Eke, Cole and Frenk,\cite{eke}
mainly with respect to $\Omega_0$. This difference seems to result from
the difference in treating the gas mass fraction as given by $\Omega_{\rm B}/
\Omega_0$ or as fixed.

\section{Cluster abundance in various models}
In this section we assume that the limiting flux $(S_\nu)_{\rm lim}$
is 6.2 mJy, corresponding to the AMIBA design, and calculate the 
limiting mass $M_{\rm lim}$ using Eq.~(\ref{eq:g13}). Then, by 
integrating Eq.~(\ref{eq:g2}), the differential number density of 
SZE clusters, $dN/(dzd\Omega)$, is obtained. 
In the following, we carry on this calculation for various models  
mentioned in \S 1, namely, the 
homogeneous models (LCDM, OCDM, SCDM and $\tau$CDM) and an 
inhomogeneous model with two homogeneous regions I and II, which 
correspond locally to OCDM and SCDM, respectively. The four 
homogeneous models correspond to $(\Omega_0, \lambda_0) = 
(0.3, 0.7), (0.3, 0), (1.0, 0)$\ and\ $(1.0, 0)$, respectively, and 
$\Gamma = 0.25$, except for SCDM with $\Gamma = 0.5$, where $\Gamma$ is
the CDM shape parameter.

\subsection{Homogeneous models}
In Figs.~1 and 2 the types of behavior exhibited by $dN/(dzd\Omega)$ in the four 
homogeneous models are shown for the two sets ($\sigma_8 (0) \Omega_0^p 
 = 0.5$ and $0.4$)  given in Eq.~(\ref{eq:g5}). Here, we use units of
deg$^{-2}$ and set $\Omega_{\rm B} h^2 = 1.70 \times 10^{-2}$.
 It is found that (1) the peaks in the low-density models are at epochs
$z = 0.17$ and $0.07$ for $\sigma_8 (0) \Omega_0^p = 0.5$ and $0.4$, 
respectively, and the number density decreases as a function of $z$
more slowly for larger $\sigma_8 (0)$, and (2) for $z = 0.2$, the number 
density in LCDM is larger by factors $>10$ than that in SCDM in both
cases of $\sigma_8 (0)$. This trend of latter behavior is the same as that
seen in Fig.~3 of the paper of Holder et al.,\cite{hold00} but it is 
different from the behavior displayed in Fig.~1 of the paper of Fan and
Chiueh.\cite{fanchiu} 
This difference comes mainly from that of the factor $\Omega_0$ used in 
the expressions of $S_\nu$ \ in the two approaches. 

Next, we consider the ratio $r$ of the number \ $[N(<0.5)]$ \ of clusters 
 with $z<0.5$ to the number \ $[N(>1.0)]$ \ of clusters with $z> 1.0$ 
in various models, following Fan and Chiueh. In Table~I, the ratios 
for OCDM and LCDM are listed, while the ratio for SCDM is omitted, because
it is very large compared with that for the other two models. It is found that the
ratio $r$ is sensitive to the value of \ $\sigma_8(0)$ \ as well as 
the model parameters. Accordingly, the observation of $r$ may yield a significant
constraint on them.  

\begin{figure}
\centerline{\includegraphics[width= 12cm]
{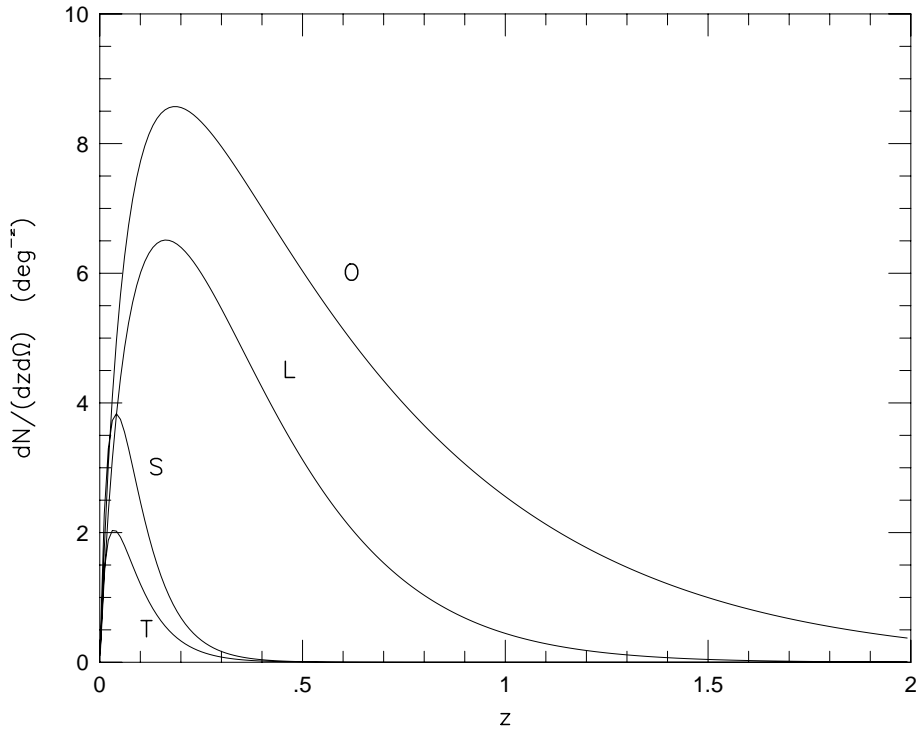}}
\caption{$dN/(dzd\Omega)$ for $\sigma_8 (0) \Omega_0^p = 0.5$ in homogeneous
models. L and O denote the models (LCDM and OCDM) with $(\Omega_0, \lambda_0) 
= (0.3,0.7)$ and $(0.3, 0)$, respectively, in which $\Gamma = 0.25$
and $h = 0.7$. S and T denote the Einstein-de Sitter models, in which 
$\Gamma = 0.5$ and $0.25$, respectively, and $h = 0.5$.
 \label{fig:1}}
\end{figure}
\begin{figure}
\centerline{\includegraphics[width= 12cm]
{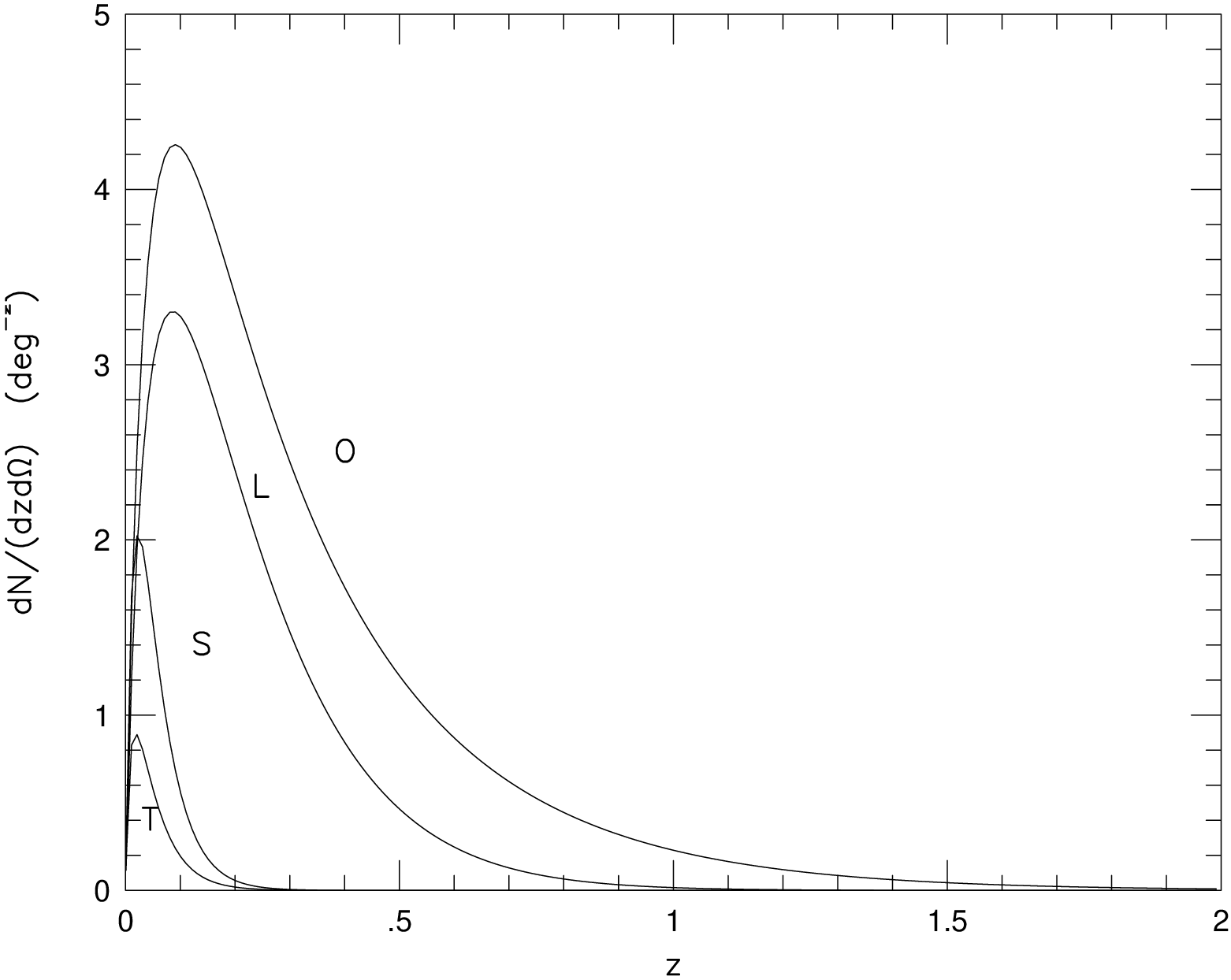}}
\caption{$dN/(dzd\Omega)$ for $\sigma_8 (0) \Omega_0^p = 0.4$ in homogeneous
models. L and O denote the models (LCDM and OCDM) with $(\Omega_0, \lambda_0) 
= (0.3,0.7)$ and $(0.3, 0)$, respectively, in which $\Gamma = 0.25$
and $h = 0.7$. S and T denote the Einstein-de Sitter models, in which 
$\Gamma = 0.5$ and $0.25$, respectively, and $h = 0.5$.
 \label{fig:2}}
\end{figure}
\begin{figure}
\centerline{\includegraphics[width= 12cm]
{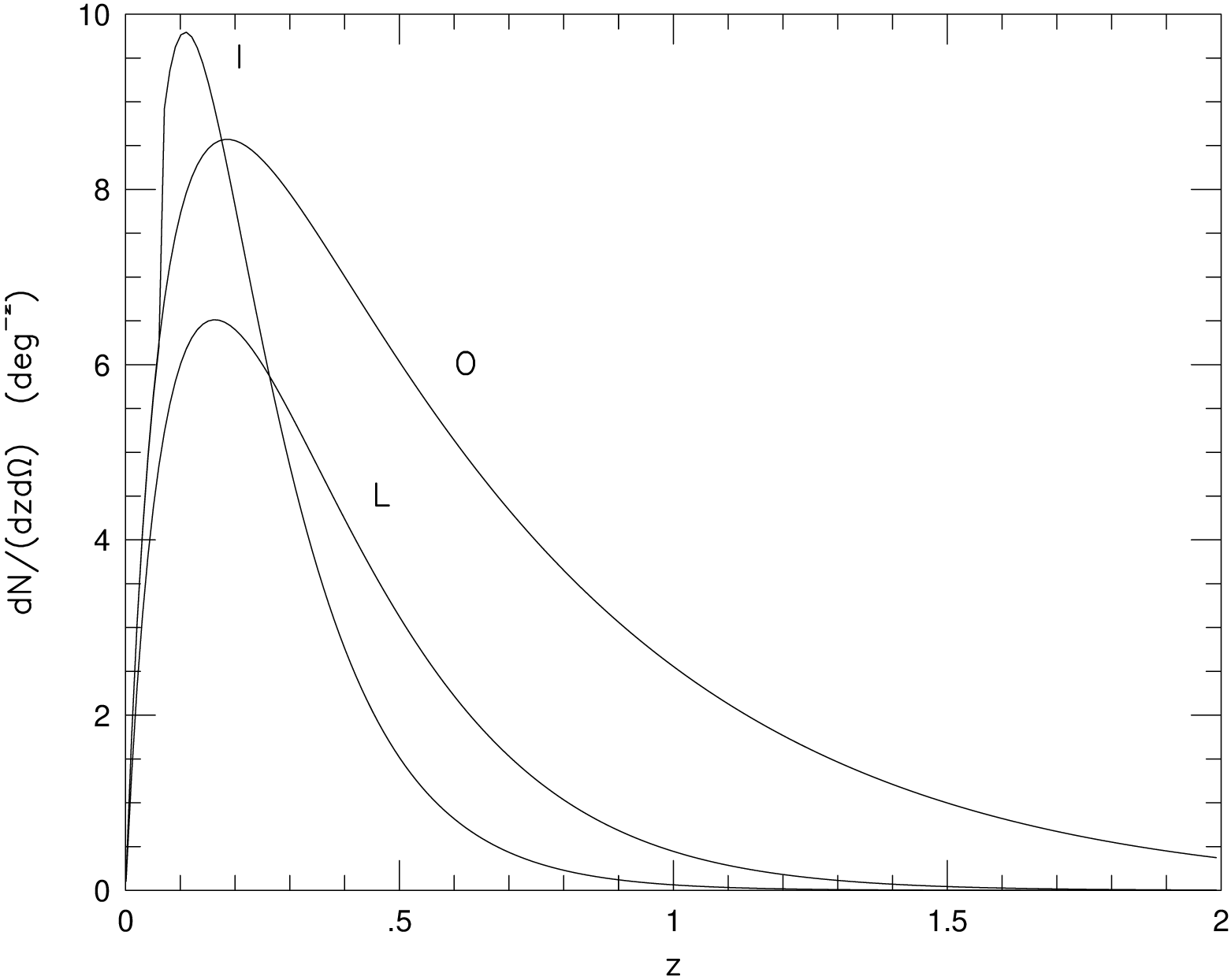}}
\caption{$dN/(dzd\Omega)$ for $\sigma_8 (0) \Omega_0^p = 0.5$ in an
inhomogeneous model (I) with $\zeta^2 = 2.0$ in comparison with the
two homogeneous models L and O.
 \label{fig:3}}
\end{figure}
\begin{figure}
\centerline{\includegraphics[width= 12cm]
{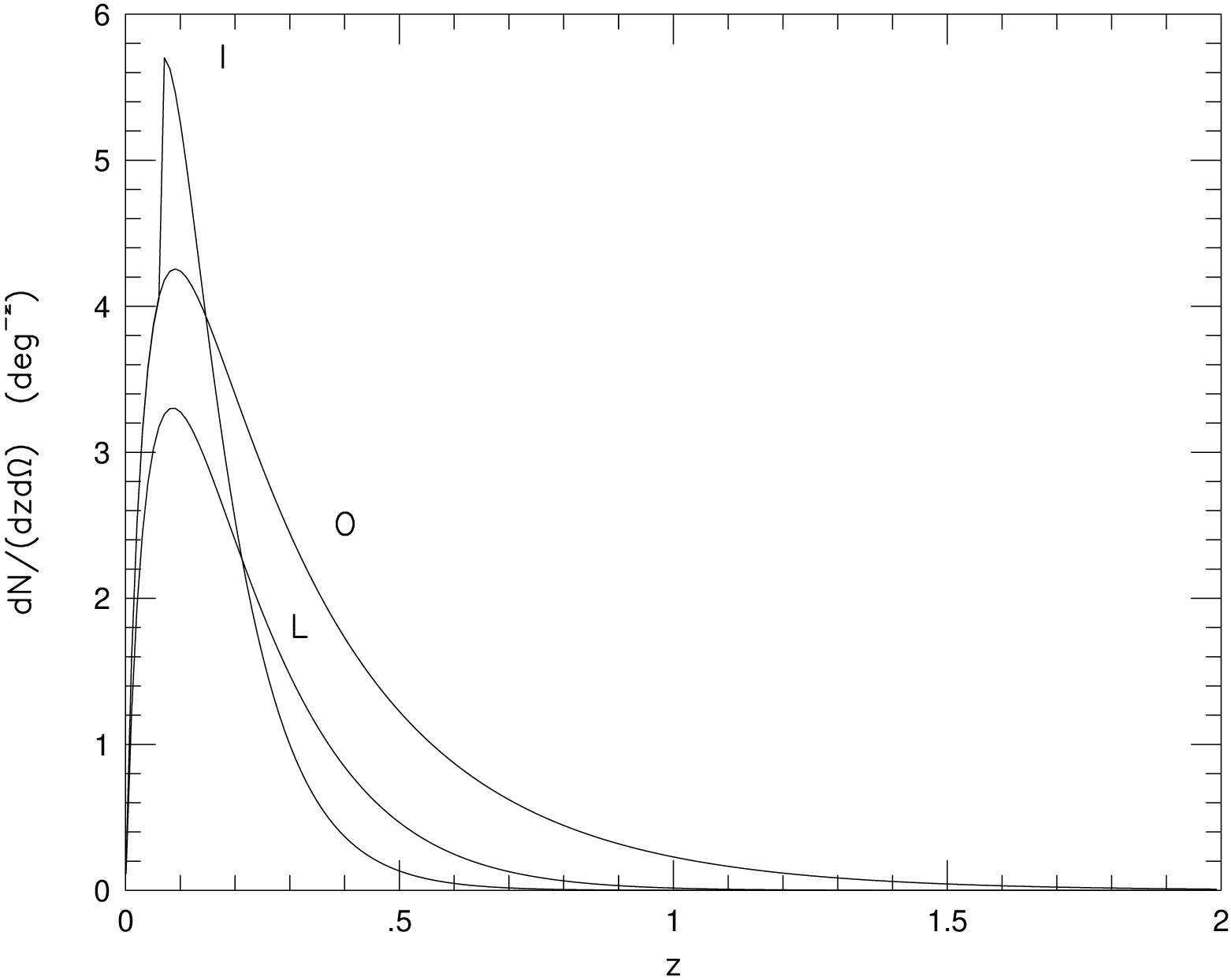}}
\caption{$dN/(dzd\Omega)$ for $\sigma_ 8 (0) \Omega_0^p = 0.4$ in an
inhomogeneous model (I) with $\zeta^2 = 2.0$ in comparison with the
two homogeneous models L and O.
 \label{fig:4}}
\end{figure}
\begin{figure}
\centerline{\includegraphics[width= 12cm]
{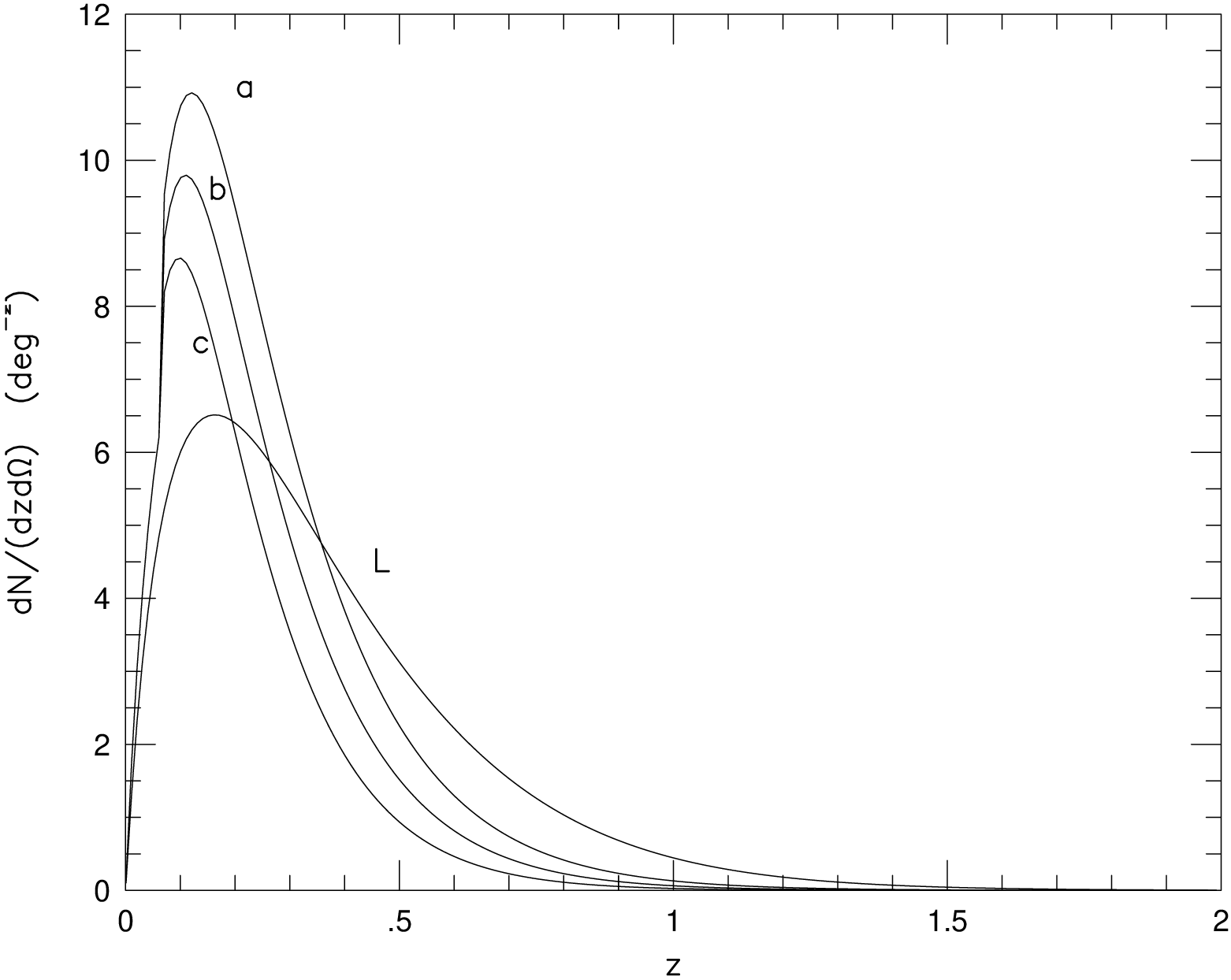}}
\caption{$dN/(dzd\Omega)$ for $\sigma_8 (0) \Omega_0^p = 0.5$ in 
inhomogeneous models with $\zeta^2 = 2.2, 2.0$ and $1.8$, which are denoted by
a, b and c, respectively.
 \label{fig:5}}
\end{figure}
\begin{figure}
\centerline{\includegraphics[width= 12cm]
{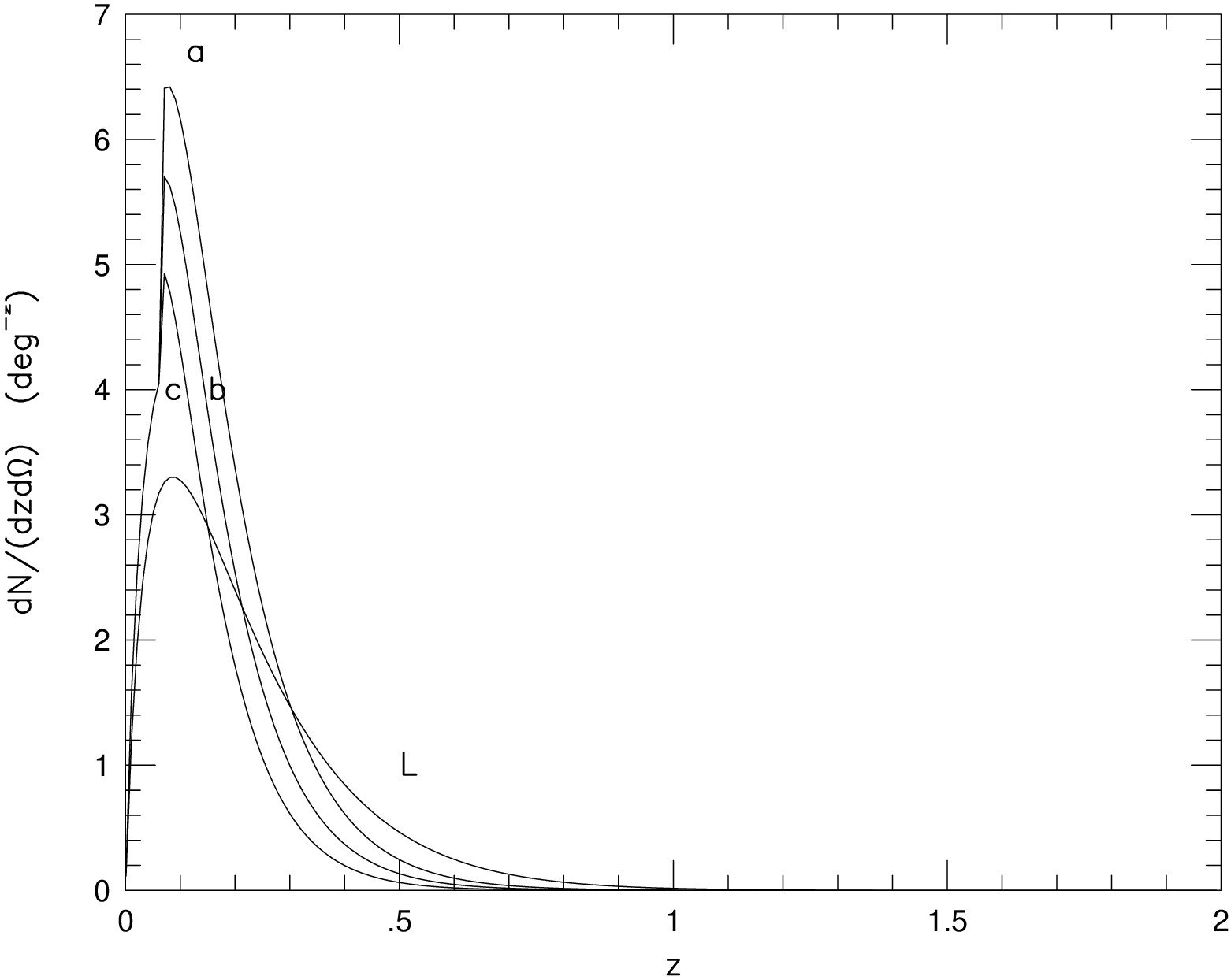}}
\caption{$dN/(dzd\Omega)$ for $\sigma_8 (0) \Omega_0^p = 0.4$ in 
inhomogeneous models with $\zeta^2 = 2.2, 2.0$ and $1.8$, which are denoted by 
a, b and c, respectively.
 \label{fig:6}}
\end{figure}
%
\begin{table}
\centering
\caption{The ratio $r \equiv N(<0.5)/N(>1.0)$ in various models 
for two values of $\sigma_8(0) (\Omega_0)^p$, where  
$p$ is \ $0.45$ \ and \ $ 0.53$ \ for models with \ $\lambda_0 = 0$ \ 
and flat models with \ $\Omega_0+\lambda_0 =1$.} 
\label{tab:1}
\begin{tabular}{|c|c|c|} \hline
{$\sigma_8(0) (\Omega_0)^p$} & models & ratio $r$ \\ \hline  
$$ & OCDM\ $(0.3, 0.0)$ & $2.67$ \\ 
$$ & LCDM\ $(0.3, 0.7)$ & $25.0$  \\ 
$0.5$ & Inhom. \ $(\zeta^2 = 2.2)$ & $141 $ \\
$$ & Inhom. \ $(\zeta^2 = 2.0)$ & $269 $ \\
$$ & Inhom. \ $(\zeta^2 = 1.8)$ & $586 $ \\
  \hline
$$ & OCDM \ $(0.3, 0.0)$ & $19.5$ \\ 
$$ & LCDM \ $(0.3, 0.7)$ & $442$  \\ 
$0.4$ & Inhom. \ $(\zeta^2 = 2.2)$ & $4.91\times 10^3 $ \\
$$ & Inhom. \ $(\zeta^2 = 2.0)$ & $1.33\times 10^4 $ \\
$$ & Inhom. \ $(\zeta^2 = 1.8)$ & $4.56\times 10^4 $ \\
  \hline
\end{tabular}
\end{table}

\subsection{Inhomogeneous cosmological models}
Here, we consider the inhomogeneous model with inner and 
outer homogeneous regions, as described in \S 1. For use
of the Press-Schechter formalism in the two regions, we assume
different spherical collapses in the two regions separately. 
Accordingly, when we use  Eq.~(\ref{eq:g13}) for  $S_\nu$,
the parameters $\Omega_0^{\rm I}$ and $\Omega_0^{\rm II}$ 
in the homogeneous models are used in the inner and outer regions, 
respectively. Because the inhomogeneity is assumed to form from an
adiabatic perturbation, the ratio of the baryon density parameter to
the total density parameter is equal in the two regions; that is, we have 
$\Omega_{\rm B}^{\rm II}/\Omega_0^{\rm II} = \Omega_{\rm B}^{\rm I}
/\Omega_0^{\rm I}$. \  This ratio has the value  $0.116$ for 
$\Omega_{\rm B}^{\rm I} {h_{\rm I}}^2 = 0.017$. In this case, we have
$\Omega_{\rm B}^{\rm II} {h_{\rm II}}^2 = 0.029$, which is consistent
with the value $\sim 0.025$ derived from the measurement of the
deuterium abundance using high-redshift QSOs\cite{deut}, while
the inner value $\Omega_{\rm B}^{\rm I} {h_{\rm I}}^2 = 0.017$ is also
consistent with the locally observed value.\cite{lith}
Moreover, the value of $\rho_{\rm vir}^i$ corresponds to $\Omega_0^i$ for 
each $i$, and for the angular-diameter distance $d_{\rm A}$, we use the 
expression in the inhomogeneous model given in previous papers.\cite{tma,tmc} 

The inner value of $\sigma_8$, $[\sigma_8(0)]^{\rm I}$, can be
determined through local observations, in the same way as in 
homogeneous models using
\begin{equation}
  \label{eq:h1}
[\sigma_8(0)]^{\rm I} (\Omega_0^{\rm I})^p = 0.5 \ {\rm or} \ 0.4.
\end{equation}
In the outer region II, on the other hand, it is to be noted that
$[\sigma_8(0)]^{\rm II}$ is larger than the local value ($[\sigma_8
(0)]_{\rm loc}$) given in the inner region by the local observations, because
the growth rate of density perturbations and the two-point correlation
functions (of clusters) in the outer region are larger than those in the 
inner region, due to the difference between the model parameters. As a result, 
 the present correlation function in the outer region also is larger 
than that in the inner region for a given same initial amplitude of density
perturbations. Here, from the definitions of $\sigma_8(0)$ and the
two-point correlation function $\xi(R)$ \ (see Suto's review
paper\cite{suto93}, for instance),\  
$\sigma_8(0)$ and $[\xi(R)]^{1/2}$ for a fixed distance
$R$ are proportional for a given same functional form of the power
spectrum. For the model parameters we use in the regions I and II, the
ratio of the correlation functions ($\equiv \zeta^2$) is found to be
of order 2, according to the theoretical analyses of Ref 32).  For the 
local value of $\sigma_8(0)$ specified  
by
\begin{equation}
  \label{eq:h2}
[\sigma_8(0)]_{\rm loc} (\Omega_0^{\rm II})^p \equiv 0.5 \ {\rm or} \ 0.4,
\end{equation}
therefore, we can adopt the value of $\sigma_8(0)$ in the region II
given by
\begin{equation}
  \label{eq:h3}
[\sigma_8(0)]^{\rm II} = \zeta [\sigma_8(0)]_{\rm loc}, 
\end{equation}
with $\zeta^2 \sim 2$. Then we have \ $[\sigma_8(0)]^{\rm I}/[\sigma_8(0)]^{\rm II}
\sim 0.3^{-0.45}/\sqrt{2} = 1.22$ for $\zeta^2 = 2$.

In Figs. 3 -- 6, the types of behavior of $dN/(dzd\Omega)$ for inhomogeneous 
models are compared with those for homogeneous models. Here we use units 
of deg$^{-2}$. In Figs. 3 and 4 (for $\sigma_8 (0) \Omega_0^p 
 = 0.5$ and $0.4$) we compare the case $\zeta^2 
= 2$ with the homogeneous models LCDM and OCDM. It is found from Figs. 1 -- 4 that 
for all $z$, the number density of clusters in the inhomogeneous 
model is much larger than that in SCDM and that for $z < 0.25 \
(0.21)$, it is larger than that in LCDM for 
$\sigma_8 (0) \Omega_0^p  = 0.5 \ (0.4)$. Also, for $z = 0.6$, it is 1/2 \ 
(1/3) of that in LCDM for 
$\sigma_8 (0) \Omega_0^p  = 0.5 \ (0.4)$. Hence, it is concluded that
SCDM can be ruled out, because of the observed existence of
clusters with $z = 0.6$ -- $0.8$. However, the present inhomogeneous model may be 
consistent with the observed existence of high-redshift clusters, like LCDM.　

In this inhomogeneous model, it is found on the other hand that for $z \sim 0.1$,
the cluster number density in the outer region is larger by a factor of $1.5$
-- $2.0$ than that in the inner region, and it is large also in comparison
with the number density in LCDM. In Figs. 5 and 6 (for $\sigma_8 (0) \Omega_0^p 
 = 0.5$ and $0.4$) the cases with $\zeta = 2.2, 2.0$ and $1.8$ are compared with
the case of LCDM. It is found that the cluster number density in the outer
region depends strongly on the value of $\zeta$, and that it decreases as 
a function of $z$ more slowly for larger $\sigma_8(0)$. 
If $\zeta$ were equal to 1.0, the cluster number density in the outer
region would reduce to that in the S model. The observation
of the cluster number density for $z \sim 0.1$, therefore, may provide a
stringent constraint on the value of $\zeta$.

The physical explanation of the behavior discussed above is basically that more
clusters form in the outer region because the growth rate of density
fluctuations there is larger than in the inner region and is characterized 
by a value of $\zeta$ larger than unity. 
 
In Table I, values of the ratio $r$ are listed for $\sigma_8 (0) \Omega_0^p  =
0.5$ and $0.4$ and for  $\zeta^2 = 2.2, 2.0$ and $1.8$.
This shows that the ratio $r$ increases with $\zeta$ and that $r$ in the 
inhomogeneous 
model with $\zeta^2 = 2.0$ is larger by a factor $\sim 10$ than $r$ in the LCDM.

\section{Concluding remarks}
In the inhomogeneous model, the cluster number density in the outer region 
is larger than that in the inner region, as shown in the previous section.
On the other hand, the galactic number densities in the two regions are nearly 
equal, as found in observational galactic surveys and studied 
theoretically in inhomogeneous models. We find that the number of galaxies 
within clusters in the outer region is smaller (by a factor of $1.5 -- 2.0$) than
that in the inner region, and therefore the mass-luminosity ratio $M/L$ in 
the outer region may be larger than that in the inner region by the same factor,
as long as $L$ of clusters is produced by galaxies within clusters. This result is 
interesting in connection with the observation of $M/L$ of clusters made by 
Bahcall and Comerford\cite{comerf} (cf. their Table I), in which we find that 
the mean value of $M/L$ for clusters with $z > 0.07$ is larger by a factor of 
$\sim 1.7$ than that for clusters with $z < 0.07$. 

For the observed values of $\sigma_8(0)$, \ the values obtained from
the cluster abundance in the region $z < 0.1$\cite{pen,eke,pierp} seem to be 
larger by a factor 
$\sim 1.5$ than the values in the region $z> 0.1$.\cite{bahc02,viana02}
The difference between $\sigma_8(0)$ in these two regions may be due to the 
uncertainties,
but if it is real, it may suggest the inhomogeneity of $\sigma_8(0)$. This is also 
interesting from the viewpoint of our inhomogeneous models, because $\sigma_8(0)$
in the inner region is larger by a factor $\sim 1.2$ than $\sigma_8(0)$ in 
the outer region. 
 
The above-described characteristic behavior of the cluster abundance in the
outer region of our inhomogeneous models will, moreover, be similarly found not
only through the Sunyaev-Zeldovich effect, but also
through X-ray and usual optical observations. A comparison
of the theoretical and observational results for the latter
methods with $2 \geq \zeta > 1$ may give some interesting and severe
constraints on the value of $\zeta$. This must be studied subsequently.   

If the observed inhomogeneity of the Hubble constant is indeed real, it cannot 
be explained by homogeneous models but only by inhomogeneous models,
 while the accelerating behavior of SN1a
can be explained also by models with a Hubble constant inhomogeneity.
Therefore, we finally add that
more accurate measurements of the Hubble constant in gravitational
lensing and SZE are desired.\cite{oguri} The data for $z = 1.7$\cite{new} 
are naturally consistent with the inhomogeneous model with $(\Omega_0^{\rm I}, 
\Omega_0^{\rm II}) = (0.3, 1.0)$,\cite{tmd} while the consistency 
with LCDM has been argued as being due to the possible contribution of 
gravitational lensing to these data.\cite{lewis,mort,benit} To clarify 
this difference, we must 
wait for the day when many data of SN1a with $z >1.5$ are obtained.

\section*{Acknowledgements}
The author thanks A. Taruya for helpful discussions on the
$\sigma_8(0) - \Omega_0$ relation and the referees for kind
suggestions for improving the manuscript. 
This work was supported by a Grant-in Aid for Scientific Research 
(No.~12440063) of the Japanese Ministry of Education, Culture, Sports, Science and 
Technology. He also acknowledges use of the YITP computer system for 
 numerical analyses.


\end{document}